\documentclass[showpacs,twocolumn,pre]{revtex4-1}%change to twocolumn

\usepackage{amsmath}
\usepackage{latexsym}
\usepackage{epsfig}
\usepackage{makeidx}
\usepackage{amsfonts}

\begin{document}

\title{Subdiffusive Transport in Heterogeneous Patchy Environments}
\author{Sergei Fedotov}
\author{Helena Stage}
\affiliation{School of Mathematics, The University of Manchester, Manchester M13 9PL, UK}

\begin{abstract}
Transport across heterogeneous, patchy environments is a ubiquitous phenomenon spanning fields of study including ecological movement, intracellular transport and regions of specialised function in a cell. These regions or patches may be highly heterogeneous in their properties, and often exhibit anomalous behaviour (resulting from e.g. crowding or viscoelastic effects) which necessitates the inclusion of non-Markovian dynamics in their study. However, many such processes are also subject to an internal self-regulating or tempering process due to concurrent competing functions being carried out. In this work we develop a model for anomalous transport across a heterogeneous, patchy environment subject to tempering. We show that in the long-time an equilibrium may be reached with constant effective transport rates between the patches. This result is qualitatively different from untempered systems where subdiffusion results in the long-time accumulation of all particles in the patch with lowest anomalous exponent, $0<\mu<1$.
\end{abstract}

\maketitle

%DEADLINE MAY 10TH

\section{Introduction}

Transport processes in heterogeneous, patchy environments is an active area of research with a multitude of applications (depending on what is meant by a patch, what is being transported and the variables affecting this transport). The notion of patches is particularly prevalent in biophysical and ecological models, describing the movement and competition of animals across different terrains or territories \cite{ecopetrovskii}, the transport of proteins between cellular organelles with different functions \cite{bressloff}, and large scale transport between tissues of different utility in an organism \cite{SergeiBook, mendezbook}. %\cite{ecopatchpnas, ecopetrovskii, ecobook}.

It is of particular importance for cellular transport given the wealth of processes which rely on transport from centrosomes or peripheral organelles to e.g. the cell membrane \cite{patchycell, patchycell2, patchycell3}. That is, where the transport taking place is dependent on local changes in function. It has previously been shown that the breakdown of these transport processes are intrinsically linked to certain diseases such as diabetes, Alzheimer's, cancer and cardiovascular problems \cite{patchycell2, disease1,disease2}.\newline

It is well-known that cell environments are crowded, leading to the trapping of particles (be they vesicles, enzymes or protein complexes), or subject to significant viscoelastic forces, inhibiting efficient transport, such that the overall transport is observed to be subdiffusive \cite{subdiff1, subdiff3, golding, motors}. Such transport is understood to occur more slowly than the Brownian equivalent and is often observed via mean square displacement (MSD) measurements of the form $\left<X^2(t)\right>\sim t^\mu$ where $0<\mu<1$ \cite{firststeps}. Models which study transport in these environments often assume that the anomalous exponent $\mu$ is constant throughout the environment. However, cells are known to be highly heterogeneous structures such that this assumption is (mostly) not a realistic approximation.

A natural consequence of the cell heterogeneity is switching between passive and active transport. The latter case, e.g. movement aided by motor proteins along microtubules or actin filaments, can lead to superdiffusive behaviour which has an MSD $\left<X^2(t)\right>\sim t^\mu$ where $1<\mu<2$ \cite{firststeps, superdiff1, superdiff2, superdiff3, superdiff4}. Both sub- and superdiffusive transport can be regarded as \textit{anomalous}. However, due to the diverging first moment of subdiffusive transport, we shall refer to these patches as being \textit{anomalously trapping}. In the superdiffusive case the transport is faster than what arises from standard Brownian motion. Here also it is usually assumed that $\mu$ is constant, which again may be an oversimplification.\newline
We are interested in the effects of allowing for heterogeneities in anomalous transport processes. Previous work on the topic of heterogeneous anomalous exponents can be found in \citep{nonlin-sokolov, nonlin-sergeisteve, nonlin-eli, nonlin-sergeinick, nonlin-sergeinick2, nonlin-other}.

The full morphology (and associated consequences for a transport process) in a cellular region is often analytically intractable, leading to the creation of simplified models which preserve the key features of the transport process in question. This is often done by the introduction of patches (regions of similar properties) and the transport of cargo between these. However, by including additional tempering effects in the patches we feel this assumption is better justified. By this local similarity assumption we shall regard patches as independent of each other.

The aim of this work is to formulate the transport equations for a patchy heterogeneous environment subject to tempering effects. That is, we consider effects such as volume filling, continued transport via other pathways out of the patch, `leakage', and so on. The main challenges in doing so for a comprehensive model is to account for three different aspects: the heterogeneity of the patches, the tempering effects which may take place in each of them, and the subdiffusive transport resulting from ageing effects in each patch. In the following section we describe the approach employed to treat these effects.

\section{Structural Density Approach}
Let us consider a system containing $\eta$ patches with different escape rates. Particles jump from one patch to another according to these rates, such that the location of a particle at a certain point in time is given by $X(t)$. So if $X(0)=4$, the particle is in fourth patch at $t=0$. In general, $X(t)$ takes values according to the subscript $i$ with values $i=1,...\eta$ depending on which patch we are in.\newline 
We assume the residence time (or age) since arrival spent by a particle in a patch $i$ at time $t$ is a random quantity given by $U_i$. Once the particle leaves the patch their age $U_i$ is reset, such that the age of any particle in the patch is independent of previous visits to the patch. Movement between patches happens with rates $\gamma_{i}\left(\tau\right)$, where
\begin{equation}
\gamma_i(\tau)=\lim_{\Delta\tau\to0^+}\left(\frac{P(\tau\leq U_i<\tau+\Delta\tau | U_i\geq\tau)}{\Delta\tau}\right)
\end{equation}
\cite{hazardrate}. Notice that these rates $\gamma_i$ depend on the residence time $\tau$ the particle has spent so far in the patch.
Taking into account the variable ages spent in a patch, we introduce the structural probability densities $\xi_{i}\left(t,\tau \right)$ which obey
\begin{equation}
\xi _{i}\left( t,\tau \right) =\frac{\partial }{\partial \tau }\Pr \left\{X(t)=i,U_i <\tau \right\}.
\end{equation}%
That is, $\xi_i(t,\tau)\Delta\tau$ gives the probability of finding particles in patch $i$ at time $t$ with residence times in the interval $(\tau,\tau+\Delta\tau)$. $\xi_i$ will thus give us the distribution of residence times in each patch as desired. The total rate of change of our probability density must balance with the escape rates from the patches, which gives the balance equations for $\xi _{i}\left( t,\tau \right) $:
\begin{equation}
\frac{\partial \xi _{i}}{\partial t}+\frac{\partial \xi _{i}}{\partial \tau }=-\gamma _{i}\left( \tau\right) \xi _{i}. 
\label{eq: start-meso}
\end{equation}%
Another interpretation of \eqref{eq: start-meso} is that changes in the current patch are purely a result of those particles which leave. Whether $\gamma_i$ increases or decreases depends on the chosen functional form.\newline
If instead of studying each individual particle, we are concerned with the aggregate of all particles, we can assume that the particles are independent of each other and study their mean. In particular, we can consider the transport in terms of the mean density of particles in a patch with a certain residence time. We call this quantity the mean \textit{structural density} $n_i(t,\tau)$ where
\begin{equation}
n_{i}\left( t,\tau \right) =N\xi _{i}\left( t,\tau \right)
\label{eq: n-def}
\end{equation}%
for a total of $N$ particles in the system. \eqref{eq: n-def} is the statement that the average number of particles in each patch is representative of the probabilities which describe their movement; a mean-field approximation has been applied to the number of particles in each patch. However, this result is independent of the total number of particles, and we may well want to work with the renormalised structural density
\begin{equation}
\rho_i(t,\tau)=\frac{n_i(t,\tau)}{N},\qquad 0<\rho_i(t,\tau)<1
\end{equation}
which describes the proportion of the whole number of particles to be found in each patch. Of course, $\rho_i(t,\tau)=\xi_i(t,\tau)$ but $\xi_i$ was introduced as a probability density for a single particle moving between patches, and $\rho_i$ is a renormalised number of particles of certain residence time in a patch resulting from a mean-field approximation. From here on out we shall use the notation $\rho_i(t,\tau)$ to stress that we are working with the mean ensemble of particles, and not just an individual one. 
In analogy to \eqref{eq: start-meso}, we can thus write
\begin{equation}
\frac{\partial \rho_{i}}{\partial t}+\frac{\partial \rho_{i}}{\partial \tau }=-\gamma _{i}\left( \tau \right) \rho_{i}. 
\label{eq: start-meso-n}
\end{equation}%
This mean structural density is in principle measurable but cumbersome to obtain experimentally for crowded biological systems where particle tracking becomes challenging. A more tractable measurement is $N_i(t)$, the (renormalised) mean number of particles in a certain patch $i$ at time $t$. Then, $1=\sum_{i=1}^\eta N_i(t)$. Note that $N_i(t)$ is simply the sum of all particles with different residence times in the patch to give
\begin{equation}
N_{i}(t)=\int_{0}^{t}\rho_{i}(t,\tau )d\tau .
\label{eq: total-dens-def}
\end{equation}%
Hence, we can decompose the particles such that $1=\sum_{i=1}^\eta \int_{0}^{t}\rho_{i}(t,\tau )d\tau$. 
%From here on out, we will consider this mean field approach instead of the structural probability densities $\xi_i$. 
Now that we have established the framework for the particles, we must specify the conditions on the system.

The particles that arrive in a patch are accounted for in the boundary conditions of zero residence time. One must also provide some initial conditions regarding the patches at $t=0$. Here we assume all patches to have initial renormalised particle distributions $\rho_i^0$ with no age, such that
\begin{equation}
\rho _{i}\left( 0,\tau \right) =\rho_i^{0}\delta \left( \tau \right).
\label{eq: init-cond}
\end{equation}%
Of course, practically one does not expect all particles to simultaneously have zero residence time, but this simplification should not affect the long-term dynamics of the the system. Similarly, the boundary conditions consider the effects of new arrivals in each of the patches. We place no limitations on which of the other patches a particle enters once leaving the current patch. The likelihood of entering another patch $j$ if currently in patch $i$ is governed by the redistribution kernel $\kappa(j|i)$. We thus obtain
\begin{equation}
\rho_i(t,0)=\sum_{j=1}^\eta\int_{0}^{t}\gamma_j\left(\tau\right) \rho_j\left(t,\tau\right)\kappa(i|j) d\tau,  
\label{eq: bound-1}
\end{equation}%
which corresponds to the statement that new arrivals $(\tau=0)$ in a patch $i$ are those particles which left the other patches and subsequently entered $i$. This formulation of the transport has the advantage that it can be generalised to the non-linear case; an easier undertaking than attempting to generalise the linear renewal equations we shall introduce in the following section. By starting from the escape rate we also allow for inclusion of effects starting from a smaller scale.

The results until now are valid for any number of patches. However, there is a wealth of evidence that cellular transport can be understood as a two-patch problem, e.g. in the spreading and proliferation of glioma cancer \cite{glioma2state}, the interaction of the motor proteins with ATP when moving cargo \cite{motor2state, motor2state2} or the associated conformational changes \cite{motors, transporters}. These patches can also be considered via their pairwise interactions as discussed in multi-stage cancer growth in \cite{cancernstate}. Other examples of the applications of such two-patch models include spiny dendrites \cite{dendritessergei, dendrites2}. For ease in following the calculations we shall now be concerned with a two-patch system ($\eta=2$) which we believe retains the essential features of larger patchy systems.

\section{Two-Patch Theory}

In this section we consider two patches with different escape rates $\gamma_i(\tau)$. We assume there are two different, and independent, processes which affect the escape rate: the residence time since arrival in the patch and the total volume capacity or transportation rate from the patch. This latter effect is assumed constant for each patch; any patch may have a small but non-zero escape constant escape rate. Then,
\begin{equation}
\gamma_i(\tau)=\beta_i(\tau)+\alpha_i,
\label{eq: gamma-decomp}
\end{equation}
is composed of ageing affects via $\beta_i$ and tempering effects (such as volume exclusion, depolymerisation rate of protein complexes, etc.) via $\alpha_i$.
Clearly, if a particle leaves one patch it must enter the other with zero residence time. We call this escape process an `event'. There are thus two types of events corresponding to entry in either patch $i=1,2$.
We introduce the mean renewal density for event of type $i$
\begin{equation}
h_{i}(t)=\rho _{i}(t,0)
\label{eq: def-h}
\end{equation}%
to denote each of these possibilities. We can use the method of characteristics to solve \eqref{eq: start-meso} where we consider the residence time $\tau(t)$ to be a function of time. That is, the time at any given point $t$ equals the time when the last renewal event happened (and a particle entered the patch) which we denote $t_0$ and the residence time $\tau$ in the patch since then:
\begin{equation}
t=t_0+\tau
\label{eq: tauoft}
\end{equation}
such that $t-\tau=t_0>0$. It is important to note that this assumes that the residence time is less than the total time that has passed. That is, at the start of our measurements we assume that all particles are newly arrived in their initial patches. The solution is given by
\begin{equation}
\rho_i(t,\tau)=\rho_i(t-\tau,0)\exp\left(-\alpha_i\tau-\int_{0}^{\tau}\beta_i[v]dv\right).
\label{eq: char-int-soln}
\end{equation} 
We notice that \eqref{eq: char-int-soln} has the form of arrivals at a time $t-\tau$ which then remain for a residence time $\tau$. This is consistent with a survival probability $P(U_i>\tau)=\Psi_i(\tau)$, which is the likelihood of remaining in the $i^{th}$ patch for a duration $\tau$ starting from the time $t$. Then, 
\begin{equation}
\Psi_i(\tau)=\exp\left(-\alpha_i\tau-\int_{0}^{\tau}\beta_i[v]dv\right),
\end{equation}
such that, using \eqref{eq: def-h}, the solution of \eqref{eq: char-int-soln} becomes
\begin{equation}
\rho_i(t,\tau)=\rho_i(t-\tau,0)\Psi_i(\tau)=h_i(t-\tau)\Psi_i(\tau).
\label{eq: char-soln}
\end{equation}
By substitution of this result into \eqref{eq: total-dens-def}, we find
\begin{equation}
N_i(t)=\int_0^t\rho_i(t-\tau,0)\Psi_i(\tau)d\tau.
\label{eq: evol-dens}
\end{equation}
If we consider each component of the escape rate separately, they have associated survival probabilities $\Phi_i^\alpha(\tau)=\exp\left(-\alpha_i\tau\right)$ and $\Phi_i^\beta(\tau)=\exp\left(-\int_0^\tau \beta_i(u)du\right)$ which are linear relationships. These individually have probability density functions (PDF) which follow $\phi_i^{\alpha}(\tau)=\alpha_i\Phi_i^\alpha(\tau)$ and similarly $\phi_i^{\beta}(\tau)=\beta_i(\tau)\Phi_i^\beta(\tau)$. This is a direct result of the relation $\phi_i(t)=-\partial\Phi_i/\partial t$. Then, one can write the total survival probability as
\begin{equation}
\Psi_i(\tau)=\Phi_i^\beta(\tau)e^{-\alpha_i\tau}.
\label{eq: totalphi}
\end{equation}
This is simply the standard survival probability one obtains for an age-dependent escape rate, with an additional tempering factor resulting from the constant escape rate. 
That is, net survival requires not leaving to due ageing in a certain time interval, and not leaving due to $\alpha_i$ contributions in that same duration.
It follows from \eqref{eq: totalphi} that the residence time PDF is given by
\begin{equation}
\psi_i(\tau)=\Psi_i(\tau)\left(\alpha_i+\beta_i(\tau)\right)=\gamma_i(\tau)\Psi_i(\tau).
\label{eq: defn-pdf}
\end{equation}
So the PDF is the survival probability apportioned by the rate at which particles leave (dependent on current residence time) and the constant small escape rate $\alpha$.
In the two-patch system all particles which leave one patch enter the other, such that the redistribution kernel
\begin{equation}
\kappa(i|j)=1-\delta_{ij}=\begin{cases}
0 \text{ if } i=j\\
1 \text{ if } i\neq j
\end{cases}.
\end{equation}
That is, all particles leaving patch $1$ enter patch $2$ and vice versa. 
We can thus apply \eqref{eq: bound-1} by multiplying \eqref{eq: char-soln} by $\gamma_i$ to obtain
\begin{equation}
\gamma_i(\tau)\rho_i(t,\tau)=h_i(t-\tau)\Psi_i(\tau)\gamma_i(\tau)=h_i(t-\tau)\psi_i(\tau),
\label{eq: whatispsi}
\end{equation}
where in the last step we have used \eqref{eq: defn-pdf}.
By integrating both sides with respect to time and using \eqref{eq: init-cond} and \eqref{eq: def-h}, we obtain the renewal equations
\begin{equation}
\begin{split}
h_{1}(t) &=\rho _{2}^{0}\left[\phi _{2}^\beta(t)+\alpha_2\Phi_2^\beta(t)\right]e^{-\alpha_2\tau}\\
 +\int_{0}^{t}&h_{2}\left(t-\tau \right)\left[\phi_2^\beta(\tau)+\alpha_2\Phi_2^\beta(\tau)\right]e^{-\alpha_2\tau}d\tau,
\end{split}
\label{eq: h1int}
\end{equation}%
\begin{equation}
\begin{split}
h_{2}(t) &=\rho _{1}^{0}\left[\phi _{1}^\beta(t)+\alpha_1\Phi_1^\beta(t)\right]e^{-\alpha_1\tau}\\
 +\int_{0}^{t}&h_{1}\left(t-\tau \right)\left[\phi_1^\beta(\tau)+\alpha_1\Phi_1^\beta(\tau)\right]e^{-\alpha_1\tau}d\tau.
\end{split}
\label{eq: h2int}
\end{equation}%
Equations \eqref{eq: h1int} and \eqref{eq: h2int} describe an alternating renewal process, and are the classical starting point in the treatment of two-patch transport processes \cite{coxmiller, renewal}. A standard approach to their solution is to apply a Laplace transformation $\mathcal{L}_t\{f(t)\}(s)=\widehat{f}(s)=\int_0^\infty e^{-st}f(t)dt$ which allows us to simplify the convolution of the two quantities. Since $\alpha_i$ is constant, this leads to
\begin{equation}
\widehat{h}_{1}(s)=\frac{\rho_{2}^{0}\widehat{\psi}_{2}\left(s\right)+\rho_{1}^{0}\widehat{\psi}_{1}\left(s\right)\widehat{\psi}_{2}\left(s\right)}{1-\widehat{\psi}_{1}\left( s\right) \widehat{\psi}_{2}\left( s\right) },
\label{eq: h1}
\end{equation}
and
\begin{equation}
\widehat{h}_{2}(s)=\frac{\rho _{1}^{0}\widehat{\psi}_{1}\left( s\right) +\rho _{2}^{0}\widehat{\psi}_{1}\left( s\right) \widehat{\psi}_{2}\left( s\right) }{1-\widehat{\psi}_{1}\left( s\right) \widehat{\psi}_{2}\left( s\right) },
\label{eq: h2}
\end{equation}%
where $\widehat{\psi}_i(s)=\widehat{\phi}_i^\beta(s+\alpha_i)+\alpha_i\widehat{\Phi}_i^\beta(s+\alpha_i)$. One reason why the renewal density approach is often applied is that since the renewal density corresponds to the density of new arrivals in the patch (see \eqref{eq: def-h}), the integral%
\begin{equation}
\int_{0}^{t}h_{i}(t)dt
\end{equation}%
gives the mean number of type $i$ events in the time interval $\left( 0,t\right)$, which is often a quantity of interest.\newline

An alternative to the renewal density is the introduction of a switching term $I_i(t)$ for each patch defined as
\begin{equation}
I_i(t)=\int_0^t\gamma_i(\tau)\rho_i(t,\tau)d\tau.
\label{eq: switch-def}
\end{equation}
This switching term can be interpreted as the renormalised flux of particles leaving a patch, where these particles can have any residence time. Hence we integrate over all values of $\tau<t$. From \eqref{eq: gamma-decomp} it follows that
\begin{equation}
\begin{split}
I_i(t)=&\alpha_iN_i(t)+\int_0^t\beta_i(\tau)\Psi_i(\tau)\rho_i(t-\tau,0)d\tau,\\
=&\int_0^t\psi_i(\tau)\rho_i(t-\tau,0)d\tau,
\end{split}
\label{eq: gen-switch}
\end{equation}
where we have used \eqref{eq: total-dens-def}, \eqref{eq: char-soln} and \eqref{eq: defn-pdf}. Taking the Laplace transform of \eqref{eq: evol-dens}, we find that $\widehat{N}_i(s)=\widehat{\rho}_i(s,0)\widehat{\Psi}_i(s)$. Further, \eqref{eq: gen-switch} in Laplace space obeys $\widehat{I}_i(s)=\widehat{\psi}_i(s)\widehat{\rho}_i(s,0)$, such that we can write
\begin{equation}
\widehat{I}_i(s)=\frac{\widehat{\psi}_i(s)}{\widehat{\Psi}_i(s)}\widehat{N}_i(s)\equiv\widehat{K}_i(s)\widehat{N}_i(s).
\label{eq: gen-lapl-switch}
\end{equation}
$K_i(t)$ is defined as above in Laplace space, such that we can write $\widehat{K}_i(s)=\alpha_i+\widehat{\phi}_i^\beta(s+\alpha_i)/\widehat{\Phi}_i^\beta(s+\alpha_i)\equiv\alpha_i+\widehat{K}_i^\beta(s+\alpha_i)$. By inversion of \eqref{eq: gen-lapl-switch}, we thus find that
\begin{equation}
I_i(t)=\alpha_iN_i(t)+\int_0^tK_i^\beta(\tau)e^{-\alpha_i\tau}N_i(t-\tau)d\tau,
\label{eq: soln-flux}
\end{equation}
where it is important that the additional term $\alpha_i$ not only leads to a separate flux, but also affects the flux resulting from the escape rate $\beta(\tau)$ via the term $e^{-\alpha_i\tau}$. So even if these escape rates are assumed to be independent of each other, they still couple in the total switching between the two patches.
By differentiating \eqref{eq: total-dens-def} and using \eqref{eq: start-meso-n} and \eqref{eq: switch-def} it follows that
\begin{equation}
\begin{split}
\frac{dN_{i}}{dt}=&\rho_i(t,t)+\int_0^t\frac{\partial \rho_i(t,\tau)}{\partial t}d\tau\\
=&\rho_i(t,0)-\int_0^t\gamma_i(\tau)\rho_i(t,\tau)d\tau\\
=&\rho_{i}(t,0)-I_i(t).
\end{split}
\label{eq: total-dens-evol}
\end{equation}
For patch $1$ we can then use \eqref{eq: bound-1} to write
\begin{equation}
\begin{split}
\frac{dN_{1}}{dt}=&I_2(t)-I_1(t),\\
=&\alpha_2N_2(t)+\int_0^tK_2^\beta(\tau)e^{-\alpha_2\tau}N_2(t-\tau)d\tau\\
-&\alpha_1N_1(t)-\int_0^tK_1^\beta(\tau)e^{-\alpha_1\tau}N_1(t-\tau)d\tau,
\end{split}
\label{eq: twostate-evol}
\end{equation}
where we have used \eqref{eq: soln-flux} in the second step. The number of particles in the other patch can be found by recalling our assumption that $N_1(t)+N_2(t)=1$. In order to obtain the final equations we must now specify the form of $K_i^\beta(t)$. In the following section we consider the simple case where tempering is not present in the dynamics of the system. These results are already known, but form a good basis of comparison with later results when tempering is included.

\section{Transport Between Patches Without Tempering}
We shall begin by considering a classical two-patch model where there are no tempering effects on the following movement. This corresponds to large systems wherein volume exclusion effects are negligible, the depolymerisation rate is negligibly small, or arriving particles in the patch are subsequently transported elsewhere in the cell leaving room for new arrivals. In any case, this corresponds to the case where $\gamma_i(\tau)=\beta_i(\tau)$. There may thus still be anomalous trapping or active transport as a result of other components of the patch properties.\newline
We remind the reader that when transport between the two patches is Markovian such that $\gamma_i(\tau)=\lambda_i$ for constant $\lambda_i$, we obtain in the long-time limit as $t\to\infty$ the result that $\widehat{\psi}_i(s)=\lambda_i/(s+\lambda_i)\simeq 1-s/\lambda_i$. This is a direct consequence of the long-time limit corresponding to the case $s\to0$ in Laplace space. This simple system is governed by the equations
\begin{equation}
\frac{dN_{1}}{dt}=\lambda_2N_2(t)-\lambda_1N_1(t),
\end{equation}
\begin{equation}
\frac{dN_{2}}{dt}=\lambda_1N_1(t)-\lambda_2N_2(t).
\end{equation}
A stationary state is thus reached where $N_1^{st}=\lambda_2/(\lambda_1+\lambda_2)$ and similarly $N_2^{st}=\lambda_1/(\lambda_1+\lambda_2)$. From the definition of the renewal measure, we know that $h_1^{st}=\lambda_2N_2^{st}$ and $h_2^{st}=\lambda_1N_1^{st}$. From the above stationary distributions, it follows that both patches have renewal density
\begin{equation}
h_{i}^{st}=\frac{\lambda_1\lambda_2}{\lambda _{1}+\lambda _{2}},
\label{eq: markov-h}
\end{equation}
which has been the subject of exhaustive research over the years \cite{coxmiller}.\newline
It is important to note that a small change in these escape rates has no qualitative impact on the long-term distribution of the particles in the patches. If both patches have the same escape rate $\lambda_1=\lambda_2=\lambda$, then we expect half of the total number of particles to be found in each patch. If one perturbs this and slightly alters the rates such that $\lambda_1<\lambda_2$, there will be a proportion of particles $\lambda_2/(\lambda_1+\lambda_2)$ and $\lambda_1/(\lambda_1+\lambda_2)$ in patches $1$ and $2$, respectively. Hence, even a patch with a high escape rate contains a non-zero number of particles. In the following we explore what happens when the transport becomes anomalous.

\subsection{Two Anomalous Patches}
It is well-established that many intracellular transport processes are not Markovian, and thus are not described by a constant escape rate \cite{superdiff2, superdiff3, superdiff4}. Instead, a measure of persistence is introduced such that the likelihood of leaving the patch decreases with the residence time of the particle in the patch. This can be modelled via the escape rates
\begin{equation}
\gamma _{i}\left( \tau \right) =\frac{\mu _{i}}{\tau _{0}+\tau },\qquad 0<\mu_i<2,
\label{eq: gamma}
\end{equation}
and is characteristic of patches which we shall call \textit{anomalous}. $\tau_0>0$ is a parameter for the time scale of the movement between patches and $\mu_i$ the constant anomalous exponents. $0<\mu<1$ corresponds to subdiffusion (anomalous trapping). If $\mu_1\neq\mu_2$ the rates at which particles leave either patch differ, and one intuitively expects to find more of the particles in the patch with a smaller escape rate. From \eqref{eq: total-dens-evol} we find that
\begin{equation}
\frac{dN_{1}}{dt}=\int_0^tK_2^\beta(\tau)N_2(t-\tau)d\tau-\int_0^tK_1^\beta(\tau)N_1(t-\tau)d\tau
\end{equation}
and
\begin{equation}
\frac{dN_{2}}{dt}=\int_0^tK_1^\beta(\tau)N_1(t-\tau)d\tau-\int_0^tK_2^\beta(\tau)N_2(t-\tau)d\tau.
\end{equation}
In this case where $\beta_i(\tau)=\mu_i/(\tau+\tau_0),\ \alpha_i=0$, we find that $\psi_i(\tau)=\mu_i\tau_0^{\mu_i}/(\tau+\tau_0)^{1+\mu_i}$ and $\Psi_i(\tau)=\tau_0^{\mu_i}/(\tau+\tau_0)^{\mu_i}$. Then, $\widehat{K}_i(s)=\widehat{K}_i^\beta(s)=\widehat{\psi}_i(s)/\widehat{\Psi}_i(s)$ cannot be inverted to obtain an expression of $K_i(t)$ for all times. Instead, we can examine the long-time limit when $t\to\infty$. From \eqref{eq: defn-pdf} and applying a Laplace transformation, we can obtain expressions for $\widehat{\psi}_i(s)$. In the long-time limit we find
\begin{equation}
\widehat{\psi}_{i}\left( s\right) \simeq\begin{cases}
1-\Gamma (1-\mu _{i})\left( \tau_{0}s\right) ^{\mu _{i}}\qquad 0<\mu_i<1\\
1-s\tau_0/(\mu_i-1)\qquad\qquad 1<\mu_i<2.
\end{cases}
\label{eq: psihat-cases}
\end{equation}%

Note that the case $1<\mu_i<2$ is qualitatively similar at long times to the PDF one obtains from a patch with a constant escape rate. However, instead of a rate we have $\lambda_i\approx (\mu_i-1)/\tau_0$. \newline

In the case when both patches are anomalous with small escape rates $\mu_1<\mu_2<1$, we are concerned with the very slow transport of particles between two anomalously trapping regions. Substituting the results from \eqref{eq: psihat-cases} into \eqref{eq: h1}-\eqref{eq: h2} we find the long-time limits of the renewal densities to be
\begin{equation}
\begin{split}
\widehat{h}_{1}(s)\simeq \widehat{h}_{2}(s)=&\frac{1}{\Gamma (1-\mu _{1})\left( \tau_{0}s\right) ^{\mu _{1}}+\Gamma (1-\mu _{2})\left( \tau _{0}s\right) ^{\mu_{2}}}\\
\simeq&\frac{1}{\Gamma (1-\mu _{1})\left( \tau_{0}s\right) ^{\mu _{1}}},
\end{split}
\label{eq: anom-lim-h}
\end{equation}%
where in the second line we have used the result that $\mu _{1}<\mu _{2}$, indicating that this patch is more trapping than patch 2. This leads to the number of switching events between the patches being entirely dictated by the anomalous exponent $\mu _{1}$; a result in stark contrast with the findings from \eqref{eq: markov-h} for the Morkovian case. Naturally, if $\mu_1=\mu_2$ \eqref{eq: anom-lim-h} yields $\widehat{h}_i(s)\simeq [2\Gamma(1-\mu_i)(\tau_0s)^{\mu_i}]^{-1}$ which is equal for both patches, but as soon as the anomalous exponents change one patch completely dominates the system. All particles will tend to be found in the patch with smallest $\mu_i$, regardless of other $\mu_i$-values. This is an important result as larger systems with more patches can be affected by minor heterogeneities found in the anomalous exponents of each patch.\newline
Note that there is nothing specific about either of these patches and the reverse effect can be obtained by reversing the relation $\mu_2<\mu_1$. The aim now is to study what occurs when the patches differ and there is only one anomalously trapping patch.

\subsection{One Anomalously Trapping Patch}
We start by noting that an anomalous patch with $\mu_i>1$ is equivalent in the long-time limit to a patch with constant escape rate as shown in \eqref{eq: psihat-cases}. The result in this limit of having two different anomalous patches (one with $\mu_1<1$ and one with $\mu_2>1$) is thus the same as that of a comparison between a trapping anomalous patch with $\mu_1<1$ and a patch with constant escape rate $\lambda_2$. By the same method as before, we find
\begin{equation}
h_{1}(t)= \frac{t^{\mu _{1}-1}}{\Gamma (1-\mu _{1})\Gamma(\mu _{1})\tau_0^{\mu_1}}
\label{eq: h-lim}
\end{equation}%
as $t\rightarrow \infty$. So in the long-time limit patch 1 is dominant if $\mu_1<\mu_2$. Note that \eqref{eq: h-lim} is independent of the escape process from patch $2$: because patch $1$ dominates the trapping of particles, even if these temporarily leave the patch before returning, the time spent in patch $2$ tends to zero and consequently the renewal between the two patches becomes effectively equivalent to a single patch renewal process. That is, the renewal process effectively describes particles entering and leaving patch $1$ with no dependence on patch $2$. Heuristically, this corresponds to letting $\lambda_2\to\infty$.\newline
These results are qualitatively sound: the more trapping patch will aggregate more of the particles. However, a perhaps surprising result is that the long-time results are independent of patch $2$.
We can write these findings in terms of the renormalised structural density of the number of particles in patch $1$, such that
\begin{equation}
\rho_{1}(t,\tau )= h_1(t-\tau )\Psi_{1}\left(\tau\right)=\frac{\Psi_{1}\left(\tau\right)t^{\mu_1-1}}{\Gamma (1-\mu _{1})\Gamma(\mu _{1})\tau_0^{\mu_1}}
\end{equation}%
as $t\to\infty$. So the (mean, renormalised) number of particles in patch $1$ at time $t$ with a certain residence time $\tau$ tends to the number of particles with entered the patch at a time $t-\tau$ (described by $h_1(t-\tau)$) and which remained there for a time $\tau$ (the probability of which is given by the survival probability $\Psi(\tau)=\int_\tau^\infty\psi(u)du$). In what follows we model this aggregation of particles in the anomalously trapping patch.

\section{Linear Anomalous Aggregation}
When the renewal density follows \eqref{eq: h-lim} we know that a non-stationary anomalous aggregation
\begin{equation}
N_{1}\left( t\right) \rightarrow 1\qquad N_{2}\left( t\right) \rightarrow 0
\label{eq: aggregation-limit}
\end{equation}%
occurs in the long-time limit as $t\rightarrow \infty$. Heuristically, this should not be surprising: if patch $1$ traps more particles and does not contain limiting factors on its size (such as a carrying capacity of the patch or volume exclusion effects) then all particles will accumulate there such that the particles which enter the patch never leave. Consequently, patch $2$ will eventually be depleted.\newline
For simplicity, let us assume that the non-trapping patch has a constant escape rate such that
\begin{equation}
\gamma_i=\begin{cases}
\frac{\mu_1}{\tau+\tau_0}\quad (\alpha_1=0)\qquad &i=1\\
\lambda_2\quad (\mu_2=0)\qquad &i=2
\end{cases}
\label{eq: linear-rates}
\end{equation}
are our escape rates. This is a reasonable assumption as we have already motivated the aggregation of particles in the anomalously trapping patch.
Applying \eqref{eq: linear-rates} to \eqref{eq: start-meso-n} we obtain
\begin{equation}
\frac{\partial \rho_{1}}{\partial t}+\frac{\partial \rho_{1}}{\partial \tau }=-\frac{\mu_1 \rho_{1}}{\tau _{0}+\tau },\qquad \mu_1<1,
\label{eq: meso-constmu}
\end{equation}%
and
\begin{equation}
\frac{\partial \rho_{2}}{\partial t}+\frac{\partial \rho_{2}}{\partial \tau }=-\lambda_2 \rho_{2}, \qquad \lambda_2>0.
\end{equation}%
Solving these equations analogously to \eqref{eq: start-meso} via the method of characteristics we find that
\begin{equation}
\rho_1(t,\tau)=\rho_1(t-\tau,0)\Psi_1(\tau),
\label{eq: meso-survival}
\end{equation}
where $\rho_1(t-\tau,0)$ is the mean (renormalised) number of newly arrived particles in the patch. These come from patch $2$ where from \eqref{eq: bound-1} we know that $\rho_1(t,0)=\int_0^t\gamma_2\rho_2(t,\tau)d\tau=\lambda_2 N_2(t)$. Similarly, for patch $2$ we find that $\rho_2(t,\tau)=\rho_2(t-\tau,0)e^{-\lambda_2\tau}$. It follows that
\begin{equation}
\rho_{1}\left( t,\tau \right) =\lambda_2 N_{2}\left( t-\tau \right) \Psi_1 (\tau),\quad t>\tau
\label{eq: meso-state1}
\end{equation}
where $\Psi_1(\tau )=\int_{\tau}^\infty\psi_1(u)du$ is the standard power-law survival function
\begin{equation}
\Psi_1 (\tau )=\left( \frac{\tau _{0}}{\tau _{0}+\tau }\right) ^{\mu_1 }.
\label{eq: simple-Psi}
\end{equation}%
We are further interested in those particles transported to the other patch. To do so, we apply a more formal definition of $h_i(t)$ than what was given in \eqref{eq: def-h}. If we define an event as a particle leaving a patch, then in Laplace space the renewal density follows \cite{coxmiller}
\begin{equation}
\widehat{h}_i(s)=\frac{\widehat{\psi}_i(s)}{1-\widehat{\psi}_i(s)}=\frac{\widehat{\psi}_i(s)}{s\widehat{\Psi}_i(s)},
\label{eq: h-fraction}
\end{equation}%
where $\widehat{\psi}_i,\ \widehat{\Psi}_i$ are the Laplace transformations of the residence time PDF and survival probability, respectively.
We can thus rewrite \eqref{eq: gen-lapl-switch} as
\begin{equation}
\widehat{I}_i(s)=\widehat{K}_i(s)\widehat{N}_i(s)=s\widehat{h}_i(s)\widehat{N}_i(s).
\end{equation}
By applying an inverse Laplace transformation we then recover a new equation for the switching term
\begin{equation}
I_{1}=\frac{d}{dt}\int_{0}^{t}h_1(t-\tau )N_{1}(\tau )d\tau,\quad I_2(t)=\lambda_2 N_2(t)
\label{eq: switch-renewal}
\end{equation}%
which can be compared with \eqref{eq: soln-flux} if desired. The switching is now entirely expressed in terms of the renewal density, and we can thus find the equations for the total number of particles in each patch. From \eqref{eq: total-dens-evol} and \eqref{eq: twostate-evol} we can use \eqref{eq: switch-renewal} to find that
\begin{equation}
\frac{dN_1}{dt}=\lambda_2 N_{2}\left( t\right)-\frac{d}{dt}\int_{0}^{t}h_1(t-\tau)N_{1}(\tau )d\tau
\end{equation}
which is valid for all time. As we have assumed a constant number of particles, $N_1(t)+N_2(t)=1$. From \eqref{eq: psihat-cases} we know that $\widehat{K}_1(s)=s^{1-\mu_1}/[\tau_0^{\mu_1}\Gamma(1-\mu_1)]$ in the long-time limit. This expression can be interpreted via the fractional derivative imposed by the Riemann-Liouville operator
\begin{equation}
_0\mathcal{D}_t^{1-\mu_i}[N_i(t)]=\frac{d}{dt}\int_0^t\frac{N_i(t-\tau)d\tau}{\Gamma(\mu_i)\tau^{1-\mu_i}}
\label{eq: rl-defn}
\end{equation}
which in Laplace space obeys $\mathcal{L}_t\{_0\mathcal{D}_t^{1-\mu_i}[N_i(t)]\}(s)= s^{1-\mu_i}\widehat{N}_i(s)$ as $s\to0$ \cite{rl-operator}. It follows that we can write
\begin{equation}
\frac{dN_1}{dt}=\lambda_2 N_{2}\left( t\right)-\frac{1}{\tau_0^{\mu_1}\Gamma(1-\mu_1)}\ _0\mathcal{D}_t^{1-\mu_1}[N_1(t)].
\end{equation}
This does not immediately provide a clearer way of understanding the movement between the two patches, but it illustrates the fractional (slow) nature of escape events from the anomalously trapping patch.
However, in the long time limit we know that aggregation of the particles will occur in patch $1$. While there may still be fluctuations in the number of particles occupying said patch, it is not unreasonable to assume it is approximately constant at larger times when the aggregation has occurred. We can then neglect the derivative $dN_{1}/dt\approx0$ such that
\begin{equation}
\lambda_2 N_{2}\left( t\right) \simeq\frac{d}{dt}\int_{0}^{t}h_1(t-\tau
)N_{1}(\tau )d\tau .
\end{equation}%
This is the statement that the rate at which particles enter patch $1$ ($\lambda_2 N_2(t)$) equals the rate at which particles leave the same patch, but which is only valid for large times. By using our assumption that the net number of particles is constant, we find
\begin{equation}
1=N_{1}\left( t\right) +\frac{1}{\lambda_2}\frac{d}{dt}\int_{0}^{t}h_1(t-\tau )N_{1}(\tau )d\tau.
\label{eq: quasi-stn}
\end{equation}%
This simple rearrangement patches that the particles are either found in patch $1$ or among those which have left patch $1$ up until now. 

In the long-time limit we can use \eqref{eq: h-fraction} to determine the behaviour of the renewal density. Using \eqref{eq: psihat-cases} the renewal density follows $\widehat{h}_1(s)=\left[(s\tau_0)^{\mu_1}\Gamma(1-\mu_1)\right]^{-1}$as indicated by \eqref{eq: anom-lim-h}. By applying an inverse Laplace transformation we obtain 
\begin{equation}
h_1(t)=\frac{t^{-1+\mu_1 }}{\Gamma (1-\mu_1 )\Gamma (\mu_1 )\tau _{0}^{\mu_1 }}
\label{eq: h-limit}
\end{equation}%
as $t\to\infty$ in analogy with \eqref{eq: h-lim}. That is, as time goes by, the number of renewal events in the patch decreases and is power law slow. This indicates that there is a slowing down in the number of particles leaving the patch. By substituting this result into \eqref{eq: quasi-stn}, we find
\begin{equation}
N_{1}\left( t\right) = 1-\frac{h_1(t)}{\lambda_2} \qquad
N_{2}\left( t\right) = \frac{h_1(t)}{\lambda_2},
\label{eq: twostate-limit}
\end{equation}%
which is consistent with the qualitative findings suggested already in \eqref{eq: aggregation-limit}. However, we now have a greater amount of detail as to how this aggregation occurs. It is important to note that while anomalous aggregation in patch $1$ is observed, we do not reach a steady-state distribution of the patch population. Movements can and do still occur between the two patches, albeit very slowly.

We shall now consider what occurs in the case when when this aggregation is tempered by an additional escape rate of the system. 

\section{Anomalous Tempering}
Having now detailed the process which occurs in the presence of anomalous patches, we shall now proceed to consider the effects of an additional tempering rate. By adding a constant $\alpha_i$ to the basic description in \eqref{eq: gamma}, we obtain
\begin{equation}
\gamma_i(\tau)=\frac{\mu _i}{\tau _{0}+\tau }+\alpha_i\qquad 0<\mu_i<1,
\label{eq: tempering-rate}
\end{equation}
where $\beta_i(\tau)=\mu_i/(\tau+\tau_0)$ is consistent with \eqref{eq: gamma-decomp}. This could change the effects observed in \eqref{eq: twostate-limit} by e.g. increasing the escape rate so as to maintain a minimum non-zero escape rate from the patch. The value (and sign) of $\alpha_i$ can be chosen according to e.g. saturation limits in the concentration of ions present in a certain transporter channel \cite{saturation} or to regulate the presence of enzymes required in protein folding \cite{saturation2}. The details and extent of this tempering are entirely determined by choices in the values of $\alpha_i$.\newline
An anomalously trapping patch with $\mu_i<1$ could thus be subject to internal regulation in the form of the tempering term which maintains the escape rate even when a large number of particles are trapped. Another possible interpretation of such a system is one wherein there is a limited binding radius beyond which arriving particles are very weakly bound, thus resulting in a constant associated escape rate. This is the simplest possible form of a self-regulating process \cite{saturation}.

As we now have two escape rates, the switching terms for these patches (defined in \eqref{eq: soln-flux}) become:
\begin{equation}
I_i(t)=\alpha_iN_i(t)+\int_0^tK_i^\beta(\tau)e^{-\alpha_i\tau}N_i(t-\tau)d\tau,
\label{eq: nonlin-switch}
\end{equation}
where we again must find the long-time limit of the integral term. Since $\widehat{\phi}_i^\beta(s)= 1-\Gamma (1-\mu _{i})\left( \tau_{0}s\right) ^{\mu _{i}}$ and $\widehat{\Phi}_i^\beta(s)=\Gamma (1-\mu _{i})\left( \tau_{0}s\right) ^{\mu _{i}}/s$, we find from \eqref{eq: gen-lapl-switch} that $\widehat{K}_i^\beta(s)= s^{1-\mu_i}/[\tau_0^{\mu_i}\Gamma(1-\mu_i)]$. The Laplace transformation of \eqref{eq: nonlin-switch} yields
\begin{equation}
\widehat{I}_i(s)=\alpha_i\widehat{N}_i(s)+\frac{(s+\alpha_i)^{1-\mu_i}}{\tau_0^{\mu_i}\Gamma(1-\mu_i)}\widehat{N}_i(s)
\end{equation}
as $s\to0$. We notice that this expression is analogous to the form of a modified Riemann-Liouville operator (see \eqref{eq: rl-defn} and \cite{rl-operator}), such that we can write
\begin{equation}
I_i(t)=\alpha_iN_i(t)+\frac{e^{-\alpha_it}}{\tau_0^{\mu_i}\Gamma(1-\mu_i)}\ _0\mathcal{D}_t^{1-\mu_i}[e^{\alpha_it}N_i(t)].
\end{equation}
This is equivalent to a tempered Riemann-Liouville operator (see e.g. \cite{temper-rl}).
By the same method as employed for \eqref{eq: total-dens-evol}-\eqref{eq: twostate-evol}, we obtain equations for the rate of change of particles in each patch:
\begin{equation}
\begin{split}
\frac{dN_{1}}{dt}=&\alpha_2N_2(t)+\frac{e^{-\alpha_2t}}{\tau_0^{\mu_2}\Gamma(1-\mu_2)}\ _0\mathcal{D}_t^{1-\mu_2}[e^{\alpha_2t}N_2(t)]\\
-&\alpha_1N_1(t)-\frac{e^{-\alpha_1t}}{\tau_0^{\mu_1}\Gamma(1-\mu_1)}\ _0\mathcal{D}_t^{1-\mu_1}[e^{\alpha_1t}N_1(t)],
\end{split}
\label{eq: full-n1}
\end{equation}
and
\begin{equation}
\begin{split}
\frac{dN_{2}}{dt}=&\alpha_1N_1(t)+\frac{e^{-\alpha_1t}}{\tau_0^{\mu_1}\Gamma(1-\mu_1)}\ _0\mathcal{D}_t^{1-\mu_1}[e^{\alpha_1t}N_1(t)]\\
-&\alpha_2N_2(t)-\frac{e^{-\alpha_2t}}{\tau_0^{\mu_2}\Gamma(1-\mu_2)}\ _0\mathcal{D}_t^{1-\mu_2}[e^{\alpha_2t}N_2(t)],
\end{split}
\label{eq: full-n2}
\end{equation}
We thus have two expressions for the number of particles entering and leaving each patch which are analogous to the results of the previous section, but which contain a faster (though still slow) transfer between the patches via the tempered Riemann-Liouville operator. However, for sufficiently large times the tempering effect introduces a cut-off in the operator and we are left with a modified constant escape rate such that
\begin{equation}
\frac{dN_{1}}{dt}=\lambda_2^*N_2(t)-\lambda_1^*N_1(t),
\label{eq: limit-constrate}
\end{equation}
where these escape rates follow
\begin{equation}
\lambda_i^*=\alpha_i+\frac{\alpha_i^{1-\mu_i}}{\tau_0^{\mu_i}\Gamma(1-\mu_i)}.
\end{equation}
Note that the tempering from $\alpha_i$ is observed in both terms, despite the initial description of $\gamma_i$ considering two independent processes. This is a result of the non-Markovian behaviour of $\beta_i(\tau)$. If the tempering effects disappear ($\alpha_i=0$) this rate $\lambda_i^*$ is no longer valid. However, when there is tempering we find that a non-zero distribution of particles can be expected across both patches. In the stationary case $dN_{1}/dt=0$, we find that
\begin{equation}
N_1^{st}= \frac{\lambda_2^*}{\lambda_1^*+\lambda_2^*},\quad N_2^{st}= \frac{\lambda_1^*}{\lambda_1^*+\lambda_2^*}
\end{equation}
which mirrors the results obtained in the case with two constant escape rates between the patches. What we conclude from this is that the presence of the tempering in $\alpha_i$ removes the anomalous effects in the long-time limit. However, over shorter time scales the anomalous aggregation effects may still dominate the dynamics.
For the simplified case of \eqref{eq: limit-constrate}, we obtain an analogous renewal density to \eqref{eq: markov-h}
\begin{equation}
h_1^{st}=h_2^{st}= \frac{\lambda_1^*\lambda_2^*}{\lambda_1^*+\lambda_2^*}.
\end{equation}

If instead of two anomalous tempered patches we consider patch $1$ to have an escape rate as given by \eqref{eq: tempering-rate}, and patch $2$ to have a constant escape rate $\lambda_2$, then \eqref{eq: full-n1} becomes
\begin{equation}
\begin{split}
\frac{dN_{1}}{dt}=\lambda_2 N_2(t)-&\alpha_1N_1(t)\\
-&\frac{e^{-\alpha_1t}}{\tau_0^{\mu_1}\Gamma(1-\mu_1)}\ _0\mathcal{D}_t^{1-\mu_1}[e^{\alpha_1t}N_1(t)].
\end{split}
\end{equation}
In this case the mean renormalised structural density from \eqref{eq: start-meso-n} follows
\begin{equation}
\frac{\partial \rho_1}{\partial t}+\frac{\partial \rho_1}{\partial \tau}=-\gamma_1(\tau)\rho_1.
\label{eq: meso-temper}
\end{equation}
Because there is tempering in effect, it is interesting to consider what occurs over longer time scales where the system has presumably equilibrated to a stationary distribution. Then, the particles are still ageing, but there is balance in the number of particles entering and leaving such that $\partial \rho_1^{st}/\partial t=0$ (note that $^{st}$ refers to any quantity in the stationary patch). \eqref{eq: meso-temper} thus becomes
\begin{equation}
\frac{\partial \rho_{1}^{st}}{\partial \tau }=-\left(\frac{\mu_1}{\tau _{0}+\tau }+\alpha_1\right)\rho_{1}^{st}.
\end{equation}
Using the fact that new arrivals in patch $1$ are given by $\rho_1(t,0)=\lambda_2 N_2(t)$, we can solve the above equation to give
\begin{equation}
\rho_1^{st}(\tau)=\lambda_2 N_2^{st}\left(\frac{\tau_{0}}{\tau_{0}+\tau }\right)^{\mu_1}e^{-\tau\alpha_1},
\end{equation}
where $N_2^{st}$ is the number of particles in patch $2$ when a steady state has been reached. Here we have identified the survival function $\Psi_1(\tau)=\tau_0^{\mu_1}e^{-\tau\alpha_1}/(\tau_{0}+\tau)^{\mu_1}$. So the number of particles with lower residence times is still high, but there is a tempering in the number of particles with long residence times. This is seen by the decaying exponential effectively `cutting off' the longer power-law tail in $\tau^{-\mu_1}$. By integration over all residence times, the mean renormalised number of particles in patch $1$ at equilibrium is given by
\begin{equation}
\begin{split}
N_{1}^{st}=&\lambda_2 N_{2}^{st}\int_{0}^{\infty }\left(\frac{\tau_{0}}{\tau_{0}+\tau }\right)^{\mu_1}e^{-\tau\alpha_1}d\tau\\
=&\lambda_2 N_{2}^{st}\tau_0^{\mu_1} e^{\tau_0\alpha_1}\alpha_1^{\mu_1-1}\Gamma(1-\mu_1,\tau_0\alpha_1),
\end{split}
\end{equation}
where $\Gamma(a,x)=\int_x^\infty t^{a-1}e^{-t}dt$ is the incomplete Gamma function \cite{rl-operator}. By definition, we know that the mean residence time spent in a patch is given by $\left<T_i\right>=\int_0^\infty\Psi_i(\tau)d\tau$. This is exactly the form we find in the above equation, such that we can write 
\begin{equation}
N_{1}^{st}=\lambda_2 N_{2}^{st}\left<T_1\right>,
\end{equation}
where $\left<T_1\right>=\tau_0^{\mu_1} e^{\tau_0\alpha_1}\alpha_1^{\mu_1-1}\Gamma(1-\mu_1,\tau_0\alpha_1)$. However, this result is only valid in the case when $\alpha_1>0$, as the integral otherwise diverges for $\mu_1<1$. Similarly for a constant escape rate we can write $\left<T_2\right>=1/\lambda_2$. Since the total number of particles is preserved, we find that
\begin{equation}
N_{1}^{st}=\frac{\left<T_1\right>}{\left<T_2\right>+\left<T_1\right>},\quad N_{2}^{st}=\frac{\left<T_2\right>}{\left<T_2\right>+\left<T_1\right>}.
\label{eq: steady-temper}
\end{equation}
Unsurprisingly, this result also mimics what we obtained when studying two patches both with constant escape rates. If we are in the long-time limit this result is equivalent to that of \eqref{eq: limit-constrate} as one can argue $\left<T_i\right>\sim1/\lambda_i$. Both patches here influence the final distribution of the particles, but we also observe that likely the particles will aggregate in the anomalous patch.

The above nicely illustrates the effects of tempering of the anomalous effects: accumulation still occurs with preference for the anomalously trapping patch, but the presence of $\alpha_i$ is such that particles could still be found in either patch. If both patches are tempered, this dominates the long-term dynamics completely (but is still a function of the anomalous exponents $\mu_i$).

\section{Discussion and Conclusion}

We have formulated the transport of particles in a heterogeneous, patchy environment and illustrated the effects of heterogeneities in the transport via the close study of transport between two patches. It has been shown that in contrast to the Markovian case (where the escape rates from patches or patches are constant), small heterogeneities in the escape rates via the anomalous exponents can lead to significant and qualitatively different distributions of particles across the system. This result remains true when constructing a larger patchy environment via the pairwise links between different patches. \newline
We have shown that a large number of particles will aggregate in the anomalous nodes (wherein the likelihood of leaving decreases with residence time), but that significant qualitative differences arise depending on whether the patch is anomalously trapping $0<\mu<1$ or not $1<\mu<2$. We have further demonstrated the effects of tempering terms in the transport processes which lead to a more even distribution of the particles than one obtains for a solely anomalously trapping patch. This is consistent with finite size effects whereupon the trapping can only occur for a binding up to a certain limit - there is thus always a minimum escape rate.

The transport of particles in cell membranes or subcellular structures are known to be subject to both non-Markovian (anomalous) transport effects as well as the ones introduced in the paper (volume exclusion, finite concentration of reaction components in producing patches, and so on).\newline
In the long-time limit particles are observed to aggregate in the anomalous (more highly trapping) patch at a power law slow rate compared to patches with a constant escape rate. This is summarised in \eqref{eq: twostate-limit}. For non-trapping anomalous patches with $1<\mu<2$, an equilibrium is reached between the two patches. The combination of these two cases with the effects resulting from tempering can be combined according to the biological system of interest.\newline
It is clear that these effects can have large implications for understanding transport mechanisms in cells. Anomalous and tempering effects are both important and significant contributions to our understanding of cell transport which in combination yield results unseen when modelling these aspects separately.

Aggregation of particles can only occur over time scales shorter than the lifetimes of the particles in question. Naturally, one expects each cell to be subject to birth-death dynamics such that certain particles may `perish' before reaching the attractive patch. One can also consider degradation rates and corruption of certain transcription processes as other inherent limits to the process. These, along with the introduction of carrying capacities in the limitations of the patches, are directions of future work to be explored.

\begin{acknowledgements}
The authors would like to thank N. Korabel and T. Waigh for fruitful discussions. This work is supported by EPSRC grant EP/N018060/1.
\end{acknowledgements}

\bibliographystyle{unsrt}
\bibliography{biblio}

\end{document}